\documentclass[reprint,twocolumn,aps,pra,floatfix,showpacs]{revtex4}%
\usepackage{amsmath}
\usepackage{amsfonts}
\usepackage{mathrsfs}
\usepackage[colorlinks,linkcolor=blue,anchorcolor=blue,citecolor=blue,urlcolor=black]%
{hyperref}
\usepackage{graphicx}
\usepackage{dcolumn}
\usepackage{bm}
\usepackage{times,epsfig}
\usepackage{amssymb}%
\begin{document}
\title{Dynamical tunneling-assisted coupling of high-Q deformed microcavities using a
free-space beam}
\author{Qi-Fan Yang}
\email{leonardoyoung@pku.edu.cn}
\author{Xue-Feng Jiang}
\author{Ya-Long Cui}
\author{Linbo Shao}
\author{Yun-Feng Xiao}
\homepage{www.phy.pku.edu.cn/~yfxiao/index.html}
\affiliation{State Key Lab for Mesoscopic Physics and School of Physics, Peking University,
P. R. China}

\begin{abstract}
We investigate the efficient free-space excitation of high-Q resonance modes
in deformed microcavities via dynamical tunneling-assisted coupling. A quantum
scattering theory is employed to study the free-space transmission properties,
and it is found that the transmission includes the contribution from (1) the
off-resonance background and (2) the on-resonance modulation, corresponding to
the absence and presence of high-Q modes, respectively. The theory predicts
asymmetric Fano-like resonances around high-Q modes in background transmission
spectra, which are in good agreement with our recent experimental results.
Dynamical tunneling across Kolmogorov-Arnol'd-Moser tori is further studied,
which plays an essential role in the Fano-like resonance. This efficient
free-space coupling holds potential advantages in simplifying experimental
condition and exciting high-Q modes in higher-index-material microcavities.

\end{abstract}

\maketitle

\section{INTRODUCTION\label{sec1}}

Over last two decades, optical whispering-gallery-mode (WGM) microresonators
(or microcavities) \cite{MicrocavitybyVahala} with high quality factors and
small mode volumes have promised lab-on-a-chip applications ranging from
fundamental physics to various photonic devices, such as nonlinear optics
\cite{PRL1986Chang,Nature2002Vahala,PRL2004Ilchenko,Comb}, cavity quantum
electrodynamics \cite{Nature2006Kimble,Nano2006Wang,XiaoPRARap}, cavity
optomechanics \cite{PRL2005Vahala,Nature2011Painter,Science2012Wang},
low-threshold microlasers
\cite{PRA1996Haroche,Microlasing,APL2003Yang,microlaserXiao} and highly
sensitive optical biosensors
\cite{APL2002Arnold,OL2006Fan,PNAS2008Keng,NatPho2010Yang,XiaoOE}. In these
applications, traditionally light is coupled into the WGM microcavities by
evanescent couplers, such as prisms \cite{PrismPRA1989}, tapered fibers
\cite{FBOL1997,PRL2000Vahala} and angle-polished fibers \cite{FbOL1999}, which
have been validated to be highly efficient. In all of these coupling
configurations, the microcavities are typically separated from the couplers by
a distance of subwavelength because the evanescent fields of WGMs extend over
a very short range. The use of the evanescent couplers, however, is not
suitable in some important applications. For example, a higher-index-material
microcavity \cite{PRL2004Ilchenko} can not be efficiently excited by the
tapered fiber due to phase mismatching. In addition, the external couplers
degrade high-Q factors (defined as $\omega\tau$ where $\omega$ denotes the
photon frequency and $\tau$ is its intracavity lifetime) in the case of the
over-coupling regime, and they are not convenient in low-temperature chambers.

It has been demonstrated that WGMs in a specially designed deformed cavity can
be directly excited by a free-space optical beam
\cite{NonresonantPumping,DynamicalAn}. This direct free-space coupling is of
importance because it is robust and requires less rigorous experimental
condition than the evanescent couplers mentioned above. This efficient
free-space coupling originates from breaking of rotational symmetry in
deformed microcavities, which produces a highly directional emission assisted
by the dynamical tunneling, different from the isotropic nature of a circular
WGM cavity \cite{Nature,DynamicalAn,NonresonantPumping}. According to the time
reversion, \textit{i.e.}, the reversibility of light path, free-space beams at
certain positions are expected to couple into the high-Q modes via chaotic
modes when they are on resonance. So far, this type of free-space coupling
technique has been demonstrated experimentally to reach a resonant efficiency
exceeding $50\%$ \cite{OE2007Wang}. A straightforward method to characterize
free-space coupling is to study its transmission property, \emph{e.g.},
transmission spectrum. In this paper, we investigate the dynamical tunneling
properties between the chaotic modes and the regular modes in detail, and
predict transmission spectra of the free-space coupling by employing a quantum
scattering theory. It is found that the spectrum can behave asymmetrically as
Fano-like lineshape \cite{Fano}, in good agreement with our recent
experimental observation \cite{onpublish}.

This paper is organized as follows. In Sec.\ \ref{sec2}, we present the
mechanism of dynamical tunneling-assisted coupling, and introduce a quantum
scattering theory to predict a general transmission in free space. It is found
that the free-space transmission spectrum includes the contribution from both
the off-resonance background and the on-resonance modulation. In
Sec.\ \ref{sec3}, we study the off-resonance background transmission in the
absence of the high-Q regular mode, corresponding to the unperturbed
scattering. The off-resonance background transmission spectrum shows periodic
modulations, which is in good agreement with both the numerical simulation and
experimental results. In Sec.\ \ref{sec4}, the on-resonance transmission
spectra are studied in detail. It is revealed that they depend strongly on (i)
the additional phase when light travels in chaos trajectories and (ii) the
rate of dynamical tunneling. Section\ \ref{phys} rigourously explains the
chaotic states and the coupling strength, with which we deduce the condition
of highest excitation efficiency. Section\ \ref{sec5} further investigates the
KAM barriers which is predominant in the dynamical tunneling process.
Section\ \ref{summary} is a short summary of the paper.

\section{Dynamical tunneling-assisted coupling\label{sec2}}

\begin{figure}[ptb]
\begin{center}
\includegraphics[width=7.5cm]{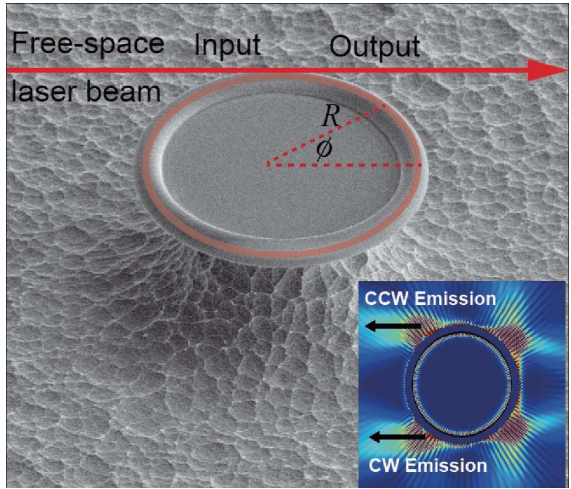}
\end{center}
\caption{(Color online) Scanning electron microscope (SEM) view of a deformed
silica microtoroid cavity. Here $R$ and $\phi$ are the polar coordinates in
the cavity plane. Red arrow denotes the laser beam. Inset: False color
illustration a resonant mode distribution obtained with wave simulation. The
two black arrows denote the dominantly directional emission toward
180${{}^{\circ}}$ far-field direction. The strength outside the cavity is
magnified for a clear show.}%
\label{Fig1}%
\end{figure}

Without loss of generality, here we consider a two-dimensional deformed cavity
made of silica with refractive index $\mathrm{n}=1.45$ as shown in
Fig.\ \ref{Fig1}, which has the boundary defined in polar coordinates as
\begin{equation}
R({\phi})=
\begin{cases}
R_{0}(1+\eta\displaystyle{\sum_{i=2,3}}a_{i}\cos^{i}\phi), & \text{for }%
\cos\phi\geq0\\
R_{0}(1+\eta\displaystyle{\sum_{i=2,3}}b_{i}\cos^{i}\phi), & \text{for }%
\cos\phi<0
\end{cases}
\end{equation}
where $R_{0}$ and $\eta$ represent size and deformation parameters,
respectively. Cavity shape parameters are set as $a_{2}=-0.1329$,
$a_{3}=0.0948$, $b_{2}=-0.0642$, $b_{3}=-0.0224$. When $\eta=1$, a highly
directional far-field universal pattern of high-Q modes has been predicted
\cite{Zou} and demonstrated experimentally \cite{Jiang}. This emission
characteristic is clearer by plotting the near-field pattern, as shown in the
inset of Fig.\ \ref{Fig1}. It can be seen that the two major emission
positions are at $\phi=\pi/2$,and $3\pi/2$, corresponding to refractive escape
from counter-clockwise (CCW) and clockwise (CW) modes, respectively. Thus, we
expect with a time reversed way, an excitation beam focused on the primary
emission position at $\phi=\pi/2$, as shown in Fig.\ \ref{Fig1}, can
eventually excite the CW resonant modes. To quantitatively study this
chaos-assisted process, we use a quantum scattering theory to model the
transmission, from which the coupling characteristic of the high-Q modes can
be obtained.

\begin{figure}[ptb]
\begin{center}
\includegraphics[width=7cm]{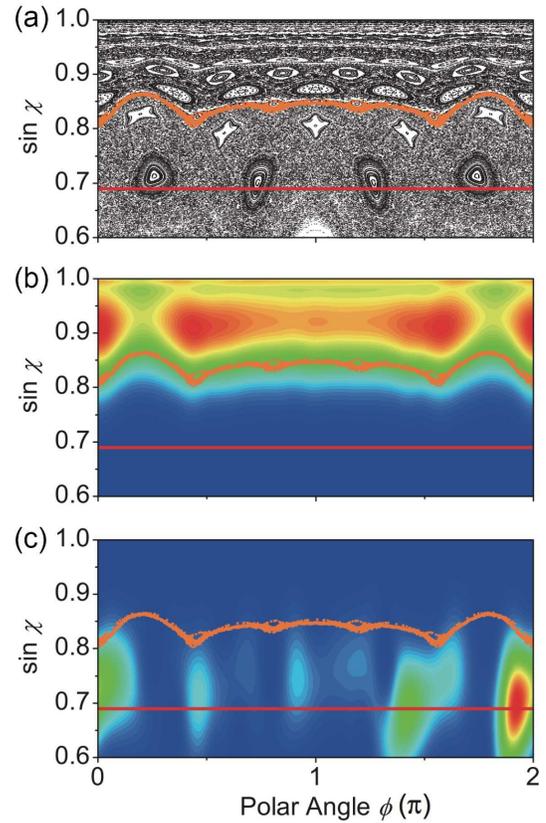}
\end{center}
\caption{(Color online) (a) A typical PSOS of the deformed microcavity. The
red solid line denotes the critical line defined as $\sin\chi=1/\mathrm{n}$.
The orange dotted line stands for a KAM torus. (b) and (c) are normalized
Husimi projection of the resonant mode and the excitation state,
respectively.}%
\label{Fig2}%
\end{figure}

Before studying the transmission spectrum, we first present the mechanism of
dynamical tunneling-assisted coupling. Poincar\'{e} surface of section (PSOS)
provides a simple and intuitive way to model the ray dynamics in deformed
microcavities by recording the angular position $\phi$ and the incident angle
$\chi$ of the rays, similar to billiards in quantum chaos. Except for an
integrable ellipse-shaped cavity, the deformed microcavity has a mixed phase
space including three types of structures: Kolmogorov-Arnol'd-Moser (KAM)
tori, islands, and chaos sea, corresponding to quasi-periodicity, periodicity,
and chaos motion of ray trajectories \cite{KAM}, as shown in Fig.\ \ref{Fig2}%
(a). KAM tori separate the PSOS into disjoint regions. As shown in
Fig.\ \ref{Fig2}(b), high-Q modes are usually localized in the regular regions
bounded by a KAM torus \cite{OL1994,Science1998,QC}. For such a localized
high-Q mode, the excitation by a free-space beam is primarily attributed to
two channels: (i) \textit{angular momentum tunneling} and (ii)
\textit{dynamical tunneling via chaos} \cite{PRAAn}. It has been demonstrated
that the dynamical tunneling dominates, since the lifetime of photons that
refract into the deformed cavity greatly increases along chaotic trajectories
\cite{NonresonantPumping}.

In the system consisting of a microcavity and unbounded medium outside, the
state $|\psi_{\omega}\rangle$ which describes the electromagnetic field
excited by the incident beam satisfies stationary Schr\"{o}dinger equation
\begin{equation}
H\left\vert \psi_{\omega}\right\rangle =\omega\left\vert \psi_{\omega
}\right\rangle , \label{eigen}%
\end{equation}
where $H$ stands for the system Hamiltonian. As mentioned above, not only
chaotic modes but the regular mode can also be excited by an appropriate
free-space beam thanks to the dynamical tunneling. This can be demonstrated by
plotting the Husimi projection \cite{Husimi} of the excitation state
$\left\vert \psi_{\omega}\right\rangle $, as shown in Fig.\ \ref{Fig2}(c).
Thus $\left\vert \psi_{\omega}\right\rangle $ can be expanded as a linear
combination of chaotic mode $\left\vert \mathrm{C}_{\omega}\right\rangle $ and
regular mode $\left\vert \mathrm{WGM}\right\rangle $ \cite{WGM}, with the
form
\begin{equation}
|\psi_{\omega}\rangle=a_{\omega}|\mathrm{WGM}\rangle+\int\mathrm{d}%
\omega^{\prime}b_{\omega}(\omega^{\prime})|\mathrm{C}_{\omega^{\prime}}%
\rangle. \label{expand}%
\end{equation}
Throughout this paper, we use regular mode and chaotic mode to describe
uncoupled states, and dynamical tunneling is the interaction between an
uncoupled regular mode and uncoupled chaotic modes. The system Hamiltonian
satisfies
\begin{subequations}
\begin{align}
\langle\mathrm{WGM}|H|\mathrm{WGM}\rangle &  =\omega_{0}-i\gamma/2,\\
\langle\mathrm{C}_{\omega^{\prime}}|H|\mathrm{C}_{\omega}\rangle &
=\omega\delta(\omega^{\prime}-\omega),\\
\langle\mathrm{C}_{\omega}|H|\mathrm{WGM}\rangle &  =V_{\omega}.
\end{align}
Here $\omega_{0}$ and $\omega$ are the frequencies of the resonant regular
mode and the incident light, respectively. The coupling coefficient between
$\left\vert \mathrm{C}_{\omega}\right\rangle $ and $\left\vert \mathrm{WGM}%
\right\rangle $, governed by the dynamical tunneling, is described by
$V_{\omega}$. The decay rate $\gamma$ consists of the intrinsic loss and the
chaos-assisted tunneling loss. In detail, the intrinsic decay rate $\gamma
_{0}$ is attributed to radiation, material absorption and scattering losses in
the cavity, while the chaos-assisted decay rate $\gamma_{1}$ describes
tunneling into the chaotic modes other than $\left\vert \mathrm{C}_{\omega
}\right\rangle $. We denote them as $\left\vert \mathrm{C}_{\omega}^{\perp
}\right\rangle $.

In this paper, we consider the chaotic modes as continuum and use a standard
quantum scattering model to interpret the transmission lineshape. Here we
assume that $|\mathrm{C}_{\omega}\rangle$ and $|\mathrm{WGM}\rangle$ are
orthogonal \cite{Tunnelingrate}. Substituting Eq. (\ref{expand}) into Eq.
(\ref{eigen}), the coefficients $a$ and $b$ are determined by
\end{subequations}
\begin{subequations}
\begin{align}
(\omega_{0}-i\gamma/2)a_{\omega}+\int\mathrm{d}\omega^{\prime}V_{\omega
^{\prime}}^{\ast}b_{\omega}(\omega^{\prime})  &  =\omega a_{\omega},\\
V_{\omega^{\prime}}a_{\omega}+\omega^{\prime}b_{\omega}(\omega^{\prime})  &
=\omega b_{\omega}(\omega^{\prime}).
\end{align}

On the one hand, applying a standard treatment \cite{Fano}, the coefficient
$b$ yields
\end{subequations}
\begin{equation}
b_{\omega}(\omega^{\prime})=[\frac{1}{\omega-\omega^{\prime}}+z_{\omega}%
\delta(\omega-\omega^{\prime})]V_{\omega^{\prime}}a_{\omega},
\end{equation}
where
\begin{equation}
z_{\omega}=\frac{\omega-\omega_{0}+i\gamma/2-F(\omega)}{|V_{\omega}|^{2}}.
\end{equation}
The shift of resonant frequency is expressed as $F(\omega)=$\textrm{v.p.}%
$\int\kappa/(2\pi(\omega-\omega^{\prime}))\mathrm{d}\omega^{\prime}$ where
v.p.\ denotes Cauchy's principle value. The reduced coupling strength $\kappa$
between $|\mathrm{C}_{\omega}\rangle$ and $|\mathrm{WGM}\rangle$ is obtained
through the Fermi's golden rule under the first Markov approximation
\cite{QuantumNoise}, with
\begin{equation}
\kappa=2\pi|V_{\omega}|^{2}. \label{kappa}%
\end{equation}
For high-Q mode in slightly deformed cavity whose intrinsic line width
$\gamma$ is orders of magnitude smaller than the resonant frequency
$\omega_{0}$, the bounds of the integral of $F(\omega)$ can be extended to
infinity, resulting in
\begin{equation}
F(\omega)=\frac{\kappa}{2\pi}\mathrm{v.p.}\int_{-\infty}^{+\infty}%
\mathrm{d}\omega^{\prime}\frac{1}{\omega-\omega^{\prime}}=0.
\end{equation}

On the other hand, the normalization condition $\langle\psi_{\omega^{\prime}%
}|\psi_{\omega}\rangle=\delta(\omega^{\prime}-\omega)$ determines the value of
$a$ by
\begin{equation}
|a_{\omega}|^{2}|V_{\omega}|^{2}[\pi^{2}+|z_{\omega}|^{2}]\delta
(\omega^{\prime}-\omega)+a_{\omega^{\prime}}^{\ast}a_{\omega}\frac{i\gamma
_{0}}{\omega^{\prime}-\omega}=\delta(\omega^{\prime}-\omega).
\label{needintegrate}%
\end{equation}
By integrating this equation over $\omega$, we have
\begin{equation}
|a_{\omega}|^{2}=\frac{1}{2\pi}\frac{\kappa}{(\omega-\omega_{0})^{2}%
+(\frac{\gamma+\kappa}{2})^{2}}. \label{aw}%
\end{equation}
In Eq. (\ref{aw}), $|a_{\omega}|^{2}$ describes the excitation probability by
the free-space beam, from which we can deduce that the FWHM (full width at
half maximum) of the regular mode is expressed as $\kappa+\gamma\equiv
\gamma_{t}$. It should be noted that $\gamma_{t}$ remains unchanged when the
free-space coupling efficiency changes \cite{DynamicalAn}, which is distinct
with the fiber taper coupling.

Finally, the transmission spectrum is calculated as
\begin{equation}%
\begin{split}
T(\omega)  &  =\left|  \langle\psi_{\omega}|S|\mathrm{in}\rangle\right|
^{2}\\
&  =|a_{\omega}|^{2}|\langle\mathrm{WGM}+\mathrm{v.p.}\int\mathrm{d}%
\omega^{\prime}\frac{V_{\omega^{\prime}}}{\omega-\omega^{\prime}}%
\mathrm{C}_{\omega^{\prime}}\\
&  +\frac{(\omega-\omega_{0}+i\gamma/2)V_{\omega}}{|V_{\omega}|^{2}}%
\mathrm{C}_{\omega}|S|\mathrm{in}\rangle| ^{2},
\end{split}
\end{equation}
where $S$ is a suitable transmission operator connecting $|\mathrm{in}\rangle$
and $|\mathrm{C}_{\omega}\rangle$, and $|\langle\mathrm{C}_{\omega
}|S|\mathrm{in}\rangle|^{2}$ describes the probability of transmitted signal
\cite{Fano}. To get a more general expression, we introduce a dimensionless
frequency detuning defined by $\epsilon\equiv(\omega-\omega_{0})/(\kappa/2)$
and the ratio $K\equiv\gamma/\kappa=(\gamma_{t}-\kappa)/\kappa$. Therefore,
the transmission is simplified to
\begin{equation}
T(\omega)=\frac{|q_{\omega}+\epsilon-iK|^{2}}{(1+K)^{2}+\epsilon^{2}}
|\langle\mathrm{C}_{\omega}|S|\mathrm{in}\rangle|^{2}. \label{T}%
\end{equation}
Here $q_{\omega}$ represents the crucial lineshape parameter of the
transmission spectrum $T(\omega)$, taking the form
\begin{equation}
q_{\omega}=\frac{\langle\varphi_{\omega}|S|\mathrm{in}\rangle}{\pi V_{\omega
}^{\ast}\langle\mathrm{C}_{\omega}|S|\mathrm{in}\rangle}, \label{q}%
\end{equation}
where $|\varphi_{\omega}\rangle=|\mathrm{WGM}\rangle+\mathrm{v.p.}%
\int\mathrm{d}\omega^{\prime}\frac{V_{\omega^{\prime}}|\mathrm{C}%
_{\omega^{\prime}}\rangle}{\omega-\omega^{\prime}}$. To give a clear
understanding, we consider two extreme cases.

(i) In classical mechanics where dynamical tunneling is forbidden, the regular
mode cannot be excited. Thus there is no interaction between the regular mode
and the chaotic mode ($\kappa\rightarrow0$), and the coefficients
$\epsilon,K\propto1/\kappa$ as well as $q_{\omega}\propto1/\sqrt{\kappa}$. In
this case the transmission spectra yields to
\begin{equation}
T_{0}(\omega)=|\langle\mathrm{C}_{\omega}|S|\mathrm{in}\rangle|,
\end{equation}
which can be regarded as the unperturbed scattering. In Sec.\ \ref{sec3} we
will discuss the unperturbed scattering, which is of much concerning about the
lineshape near resonance.

(ii) When the intrinsic loss of regular mode is negligible and the regular
mode is fully excited by a phase conjugation wave of its emission pattern,
\textit{i.e.}, $\gamma\ll\kappa$ and $K\rightarrow0$ (or namely, `complete
excitation', which will be discussed specifically in Sec. \ref{phys}), the
transmission yields a standard Fano resonance
\begin{equation}
T(\omega)=\frac{|q_{\omega}+\epsilon|^{2}}{1+\epsilon^{2}}|\langle
\mathrm{C}_{\omega}|S|\mathrm{in}\rangle|^{2}.
\end{equation}

\section{Off-resonance transmission\label{sec3}}

\begin{figure}[ptb]
\begin{center}
\includegraphics[width=7cm]{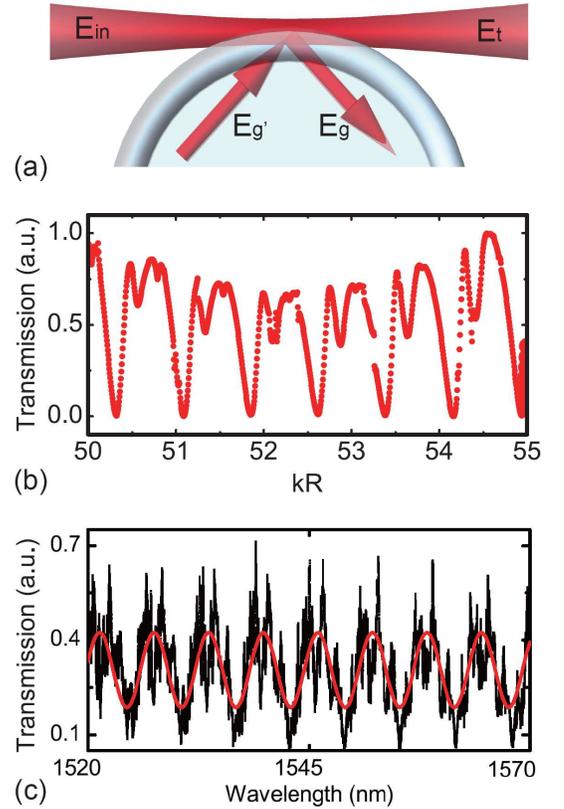}
\end{center}
\caption{(Color online) (a) The schema of the scattering process described by
Eq.\ \ref{scattering}. (b) Stimulated transmission spectrum of the free-space
excitation process obtained by boundary element method. (c) Experimental
spectrum (black) and the fitted oscillations (red).}%
\label{Fig3}%
\end{figure}

We now investigate the background scattering in the absence of the high-Q
regular mode. It has been reported that non-resonant pumping in deformed
microcavity can be well modeled by ray dynamics
\cite{Fresnel,TransmissionKorea}. In our case, the unperturbed transmission is
studied in wave optics, and it results from the interference between two
components, according to the schema shown in Fig.\ \ref{Fig3}(a): (i) the
direct transmitted amplitude $t$, and (ii) the dissipated amplitude $r$ via
diffusing inside the cavity. To give a clear picture of the interference, we
apply transmission matrix, and the amplitudes are indicated by
\begin{equation}%
\begin{pmatrix}
E_{\mathrm{t}}\\
E_{\mathrm{g}}%
\end{pmatrix}
\ =%
\begin{pmatrix}
t & g^{\prime}\\
g & t^{\prime}%
\end{pmatrix}%
\begin{pmatrix}
E_{\mathrm{in}}\\
E_{\mathrm{g^{\prime}}}%
\end{pmatrix}
. \label{scattering}%
\end{equation}
The intracavity fields $E_{\mathrm{g^{\prime}}}$ and $E_{\mathrm{g}}$ are
related by
\begin{equation}
E_{\mathrm{g^{\prime}}}=\alpha E_{\mathrm{g}},
\end{equation}
where $\alpha$ is a coefficient including the loss and the phase change in a
round trip. The transition matrix element is then given by
\begin{equation}%
\begin{split}
\langle\mathrm{C}_{\omega}|S|\mathrm{in}\rangle=\frac{E_{\mathrm{t}}%
}{E_{\mathrm{in}}}  &  =t+\frac{\alpha g^{\prime}g}{1-\alpha t^{\prime}}\\
&  =t+re^{i\theta}.
\end{split}
\label{transition}%
\end{equation}
Here $r$ and $\theta$ can be understood as the equivalent amplitude and phase
difference of forward-emitted field from the cavity, respectively. Hence, the
unperturbed transmission takes the form
\begin{equation}
T_{0}=|\langle\mathrm{C}_{\omega}|S|\mathrm{in}\rangle|^{2}=r^{2}\left(
1+(\frac{t}{r})^{2}+2\frac{t}{r}\cos\theta\right)  . \label{background}%
\end{equation}
In a wide frequency width the phase difference $\theta$ can be simplified as
$\mathrm{nk}L_{\mathrm{eff}}$ with $L_{\mathrm{eff}}$ representing the
equivalent chaotic path length of the light inside the cavity. Thus the
transmission spectrum shows periodic modulations in good agreement with
numerically simulated transmission, as shown in Fig.\ \ref{Fig3}(b). In
experiment, we focus the incident beam on the periphery of a deformed
microtoroid with the principle radius $45$ $\mathrm{\mu m}$, and the waist of
the beam is about $3$ $\mathrm{\mu m}$ \cite{Jiang}. Figure\ \ref{Fig3}(c)
reveals that the experimentally detected transmission also oscillates
periodically. Note that the narrow fluctuations in the transmission are due to
the Fabry-Perot oscillations between two lens. From experimental results the
ratio $t/r$ can be obtained by fitting the large scale transmission as shown
in red in Fig.\ \ref{Fig3}(c).

\section{On-resonance transmission\label{sec4}}

\begin{figure}[ptb]
\begin{center}
\includegraphics[width=7.5cm]{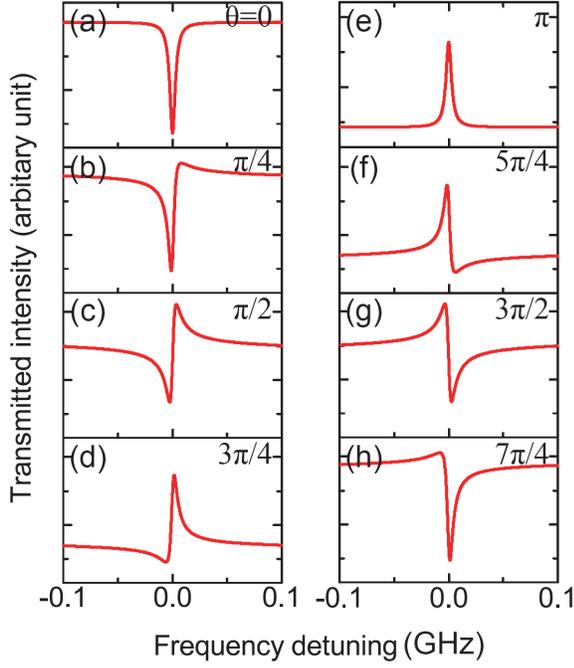}
\end{center}
\caption{(a)-(h) Calculated transmission spectra with $\kappa/2\pi,\gamma
/2\pi=0.003$ GHz and $t/r=0.2$. The phase shift between the two amplitudes
varies from 0 to $7\pi/4$.}%
\label{Fig4}%
\end{figure}

We turn to the study of the on-resonance transmission. It is noted that the
direct excitation probability of high-Q regular modes via evanescent field is
negligible due to angular momentum mismatch, so that the amplitude of
$\langle\mathrm{WGM}|S|\mathrm{in}\rangle$ in Eq. (\ref{q}) has minor
contribution of the transmission. In this case, the lineshape parameter $q$
can be reduced to a simplified form. Substituting Eq. (\ref{transition}) into
Eq. (\ref{q}), we obtain the expression of the lineshape parameter
\begin{equation}
q_{\omega}=\frac{\mathrm{v.p.}\int\mathrm{d}\omega^{\prime}\frac{1}%
{\omega-\omega^{\prime}}V_{\omega^{\prime}}^{\ast}\langle\mathrm{C}%
_{\omega^{\prime}}|S|\mathrm{in}\rangle}{\pi V_{\omega}^{\ast}\langle
\mathrm{C}_{\omega}|S|\mathrm{in}\rangle}=-\frac{ie^{i\theta}}{t/r+e^{i\theta
}}, \label{gaussianq}%
\end{equation}
where $\theta=\mathrm{nk}L_{\mathrm{eff}}$ as mentioned above. Substituting
Eq. (\ref{gaussianq}) into Eq. (\ref{T}), the on-resonance transmission can be
deduced. In the following, we will show that the lineshape of the transmission
spectrum is determined by $q_{\omega}$, which primarily depends on $\theta$,
while the modulation depth relies on the relative coupling strength described
by $K$.

Figures \ref{Fig4}(a)-\ref{Fig4}(h) plot calculated transmission spectra
against the phase difference $\theta$, which experience symmetric Lorentz
absorption dips, asymmetric Fano-like lineshapes and symmetric
electromagnetically-induced-transparency (EIT)-like peaks as $\theta$ varies
from $0$ to $2\pi$. From Fig.\ \ref{Fig3}(a), the transmission is a result of
interference between two components: the direct transmitted light and the
emitted light from the cavity, and $\theta$ actually describes the phase
difference between them. Interestingly, the on-resonance transmission appears
a symmetric dip on the background where the two components constructively
interfere ($\theta=0$ in Fig.\ \ref{Fig4}(a)), while it switches to an
EIT-like peak when they destructively interfere ($\theta=\pi$ in
Fig.\ \ref{Fig4}(e)). This is because when on resonance, the chaotic modes
refractively excited by the incident beam can couple to the regular mode via
dynamical tunneling, which results in a phase shift as energy couples back to
the chaotic modes. Hence, although the background components constructively
(destructively) interfere, such counteraction adds a destructive modulation to
the chaotic modes, reflecting a dip (peak) on the transmission.

\begin{figure}[ptb]
\begin{center}
\includegraphics[width=7.8cm]{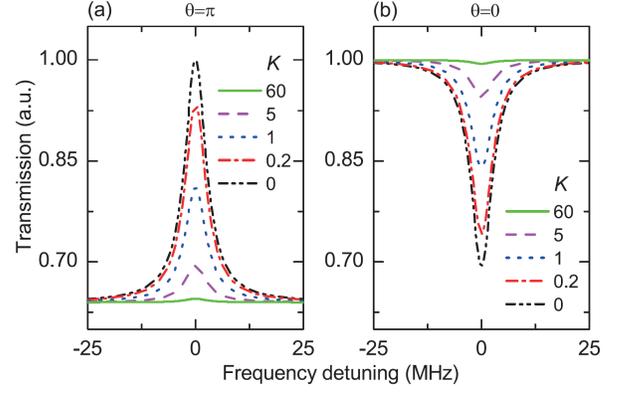}
\end{center}
\caption{(Color online) Calculated transmission spectra with $\gamma_{t}%
/2\pi=0.006$ GHz and $t/r=0.2$, and with $K$ varying from 0 to 60, for (a)
EIT-like lineshape and (b) Lorentz-like lineshape. The background is set to
the same level for each graph.}%
\label{Fig5}%
\end{figure}

As discussed above, the Fano-resonance transmission spectra can be regarded as
the modulation of the high-Q mode on the off-resonance background. Such
modulation depends strongly on the coupling strength $\kappa$ between the
chaotic modes and the regular mode according to Eq. (\ref{T}). Here we study
the two special cases: EIT-like lineshapes and Lorentz dips. As shown in the
solid curve in Fig.\ \ref{Fig5}(a), the modulation of the regular mode to the
transmission spectrum is minor when $K=60$. In this case, the excitation
probability is extremely low. As $K$ decreases (\textit{i.e.}, the dynamical
tunneling is enhanced), the height of the EIT peak increases monotonically,
where the off-resonance backgrounds are lifted to the same level. When the
loss described by $\gamma_{0}+\gamma_{1}$ is negligible compared with the
coupling strength $\kappa$, \textit{\i.e.}, $K\rightarrow0$, the EIT peak
reaches its maximum. Similarly, Fig.\ \ref{Fig5}(b) shows that the
dynamical-tunneling-induced dips become more obvious by enhancing the
tunneling, as expected.

\section{Physical meaning of the chaotic mode and `complete
excitation'\label{phys}}

At the beginning of Sec.\ \ref{sec2}, we have presented the chaotic mode
$\left\vert \mathrm{C}_{\omega}\right\rangle $. In this section, we will
further investigate the meaning of the chaotic mode and the coupling strength
to a regular mode $|\mathrm{WGM}\rangle$. To study this case in a general way,
we expand the chaotic mode as a linear combination of an orthogonal set at a
certain frequency. Using $\left\vert \mathrm{C}_{\omega(n)}\right\rangle $ to
represent the normalized $n$-th orthogonal mode, we have
\begin{equation}
|\mathrm{C}_{\omega}\rangle=\frac{1}{\sqrt{\displaystyle{\sum_{n}}\alpha
_{(n)}^{2}}}\displaystyle{\sum_{n}}\alpha_{(n)}|\mathrm{C}_{\omega(n)}\rangle,
\end{equation}
where $\alpha_{(n)}$ stands for the corresponding weight. From the coupled
mode theory, we obtain
\begin{equation}
\dot{\xi}_{r}=\displaystyle\sum_{n}g_{(n)}\xi_{(n)}. \label{cp1}%
\end{equation}
Here $\xi_{n}$ and $\xi_{r}$ represent the electric field of $|\mathrm{C}%
_{\omega(n)}\rangle$ and $|\mathrm{WGM}\rangle$, respectively, with $g_{n}$
being the coupling strength between them. Thus, the equivalent coupling
strength $|V_{\omega}|$ is derived as
\begin{equation}
|V_{\omega}|=\frac{1}{\sqrt{\displaystyle{\sum_{n}}\alpha_{(n)}^{2}}%
}\displaystyle{\sum_{n}}\alpha_{(n)}g_{(n)}.
\end{equation}
Then the reduced coupling strength between $|\mathrm{C}_{\omega}\rangle$ and
$|\mathrm{WGM}\rangle$ can be obtained as $\kappa=2\pi|V_{\omega}|^{2}$,
according to Eq. (\ref{kappa}).

Once the high-Q regular mode $|\mathrm{WGM}\rangle$\ is excited, it can
dynamically tunnel into all the chaotic modes including both $\left\vert
\mathrm{C}_{\omega}\right\rangle $ and $\left\vert \mathrm{C}_{\omega}^{\perp
}\right\rangle $. Since the chaotic modes are continuum, with first Markov
approximation, this tunneling can be considered as a spontaneous decay process
of the regular mode, described by the coefficient%
\begin{equation}
\gamma_{\omega}=\sqrt{\displaystyle{\sum_{n}}g_{(n)}^{2}}.
\end{equation}
Thus the decay rate into $\left\vert \mathrm{C}_{\omega}^{\perp}\right\rangle
$ is
\begin{equation}
\frac{\gamma_{\mathrm{1}}}{2\pi}=\left\vert \gamma_{\omega}\right\vert
^{2}-\frac{\kappa}{2\pi}=\frac{\displaystyle{\sum_{n}}\alpha_{(n)}%
^{2}\displaystyle{\sum_{n}}g_{(n)}^{2}-(\displaystyle{\sum_{n}}\alpha
_{(n)}g_{(n)})^{2}}{\displaystyle{\sum_{n}}\alpha_{(n)}^{2}}.
\end{equation}
Hence, to optimize the free-space coupling efficiency, according to Cauchy
inequality, the coefficients $\alpha_{(n)}$ satisfies
\begin{equation}
\frac{\alpha_{(1)}}{g_{(1)}}=\frac{\alpha_{(2)}}{g_{(2)}}=\cdots=\frac
{\alpha_{(n)}}{g_{(n)}}.
\end{equation}
Under this condition, it is found $\gamma_{\mathrm{1}}=0$. Neglecting the
intrinsic loss $\gamma_{\mathrm{0}}$ induced by scattering and material
absorption, we have $\kappa=\gamma_{\mathrm{t}}$, which means the incident
light is exactly a time-reversed way of the emission light from the regular
mode $|\mathrm{WGM}\rangle$. It is also in agreement with the second extreme
case discussed in Sec. \ref{sec2}, that the `complete excitation' condition
requires the incident light as the phase conjugation wave of the emission pattern.

\section{KAM barriers\label{sec5}}

Finally, we discuss how the KAM tori, behaving as barriers
\cite{PRL1986,PRL2007}, can result in a phase shift in dynamical tunneling.
This phase shift is crucial to give rise to the Fano resonance. As shown in
Fig.\ \ref{Fig2}(a), KAM tori separate the phase space into disconnected
regions, between which transport is forbidden classically, but permitted in
quantum mechanics \cite{TurnstileAn,PRLsong}, known as dynamical tunneling. To
evaluate the barrier effect, we study the potential of orbits in the PSOS. For
the sake of analytical expressions but without loss of the physics, we
investigate the orbits in a circular microcavity. The wave function of a WGM
with angular momentum number $\mathrm{m}$ in circular cavities takes the form
\begin{equation}
\Psi(\mathbf{r},\phi)=f_{\mathrm{m}}(\mathbf{r})e^{i\mathrm{m}\phi},
\end{equation}
where $f_{\mathrm{m}}(\mathbf{r})$ satisfies the radial wave equation
\begin{equation}
\lbrack-\nabla_{\mathbf{r}}^{2}-\frac{(\mathrm{n}^{2}(\mathbf{r})-1)\omega
^{2}}{c^{2}}+\frac{\mathrm{m}^{2}}{r^{2}}]f_{\mathrm{m}}(\mathbf{r}%
)=\frac{\omega^{2}}{c^{2}}f_{\mathrm{m}}(\mathbf{r}).
\end{equation}
Based on the stationary Schr\"{o}dinger equation
\begin{equation}
(-\frac{\hbar^{2}}{2\mu}\nabla^{2}+V)f_{\mathrm{m}}(\mathbf{r})=Ef_{\mathrm{m}%
}(\mathbf{r})
\end{equation}
and substituted with $E=\hbar\omega$, we deduce the effective potential
corresponding to the angular momentum number $\mathrm{m}$
\begin{equation}
V=\frac{\hbar^{2}}{2\mu}[-\frac{(\mathrm{n}(\mathbf{r})^{2}-1)\omega^{2}%
}{c^{2}}+\frac{\mathrm{m}^{2}}{r^{2}}].
\end{equation}
\begin{figure}[ptb]
\begin{center}
\includegraphics[width=7cm]{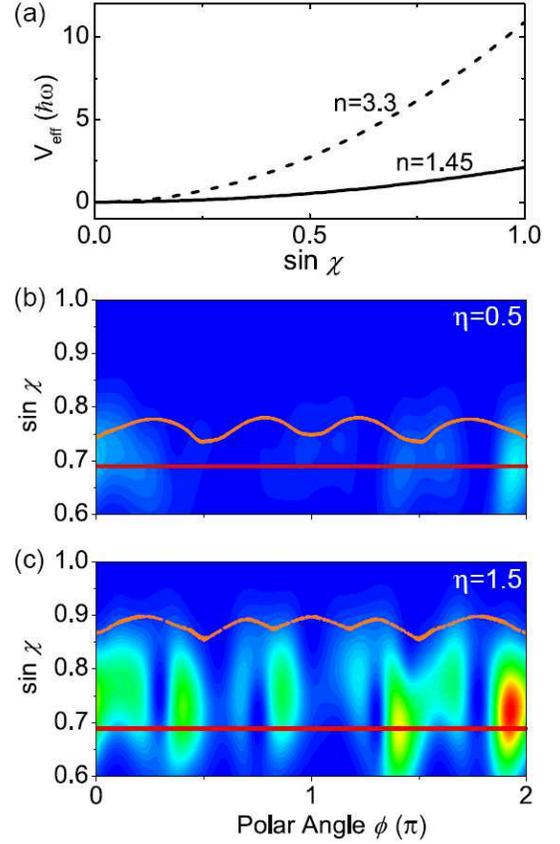}
\end{center}
\caption{(Color online) (a) Effective potential $V_{\mathrm{eff}}$ against
$\sin\chi$. The solid and the dashed curves correspond to the cases of silica
(n=1.45) and GaAs (n=3.3) microcavities, respectively. (b)-(c) Husimi
projections of the excitation state inside the cavity at non-resonant
frequency with the deformation parameter $\eta$ of the cavity setting as $0.5$
and $1.5$, respectively. These two figures are plotted in the same scale.
Orange dotted curves and red solid lines denote KAM tori and
critical-refraction lines.}%
\label{Fig6}%
\end{figure}

Extending $\mathrm{m}$ to the range of positive real numbers, from the
classical relation $\mathrm{m}=\mathrm{nkr}\sin\chi$ and the non-relativity
approximation $\mu=\hbar\omega/2c^{2}$, the effective potential of the orbits
takes the form
\begin{equation}
V_{\mathrm{eff}}(\sin\chi)=\mathrm{n^{2}}\hbar\omega\sin^{2}\chi,\label{V}%
\end{equation}
The effective potentials for different meterials (silica and GaAs) are plotted
in Fig.\ \ref{Fig6}(a). Thus, an photon at critical refraction line $\sin
\chi=1/\mathrm{n}$ has the same potential as a free-space photon, in agreement
with the Fresnel's Law. The difference in the potentials of various orbits
leads to tunneling.

In deformed microcavities, KAM tori are the residue of these invariant orbits,
and they perform as barriers in quantum mechanics. By plotting the Humisi
projection of cavities of different deformations shown in Figs.\ \ref{Fig6}%
(b)-(c), we can find that the transportation to high-Q modes are governed by
the KAM tori. Our simulation also reveals that more light can be localized in
the cavity with a larger deformation, thus leading a higher probability to
couple into regular high-Q modes.

\section{summary\label{summary}}

In conclusion, we have presented the dynamical tunneling-assisted coupling
mechanism to interpret how a free-space laser beam excites the high-Q modes in
deformed microcavities. The deformed microcavity has a mixed phase space,
where the high-Q regular modes lie in regular regions. Lifetime of photons
refracting into the cavity increases due to chaotic trajectories, which
contributes to the enhanced excitation of regular modes via
\emph{chaos-assisted dynamical tunneling}. A quantum scattering theory is
employed to describe the picture and to obtain the free-space transmission
spectra. Unlike evanescent coupling with a waveguide where the transmission
spectra behave symmetrically, this model predicts three types of transmission,
\textit{i.e.}, asymmetric Fano-like, symmetric EIT-like and Lorentz dip
lineshapes, depending on the phase difference related to the fluctuation of
background transmission. It is found that the Fano resonance is attributed to
the phase shift occurring in the dynamical tunneling into classical-forbidden
regions. Our results provide a general method to evaluate the coupling
strength between the chaos and the regular mode from the transmission spectra,
which can be further extended to the quantitative study of the dynamical
tunneling process. The efficient chaos-assisted free-space coupling is of
importance to simplifying experimental condition and exciting high-Q modes in
higher-index-material microcavities.

\begin{acknowledgments}
Q.F.Y. and Y.F.X. thank Kerry Vahala and Fang-Jie Shu for insightful
discussions. This work was supported by the 973 program (Grant No.
2013CB328704), the NSFC (Grants No. 11222440, No. 11004003, No. 11023003 and
No. 1121091), the RFDPH (Grant No. 20120001110068) and Beijing Natural Science
Foundation Program (Grant No. 4132058). Q.F.Y. and Y.L.C. were supported by
the National Fund for Fostering Talents of Basic Science (No. J1030310 and No.
J1103205), and the Undergraduate Research Fund of Education Foundation of
Peking University.
\end{acknowledgments}


\begin{thebibliography}{99}                                                                                               %


\bibitem {MicrocavitybyVahala}\emph{Optical Microcavities}, edited by K.
Vahala (World Scientific, Singapore, 2005).

\bibitem {PRL1986Chang}S.-X. Qian and R. K. Chang, \textrm{Phys. Rev. Lett.}
\textbf{56}, 926 (1986).

\bibitem {Nature2002Vahala}S. M. Spillane, T. J. Kippenberg, and K. J. Vahala,
\textrm{Nature} \textbf{415}, 623 (2002).

\bibitem {PRL2004Ilchenko}V. S. Ilchenko, A. A. Savchenkov, A. B. Matsko, and
L. Maleki, \textrm{Phys. Rev. Lett.} \textbf{92}, 043903 (2004).

\bibitem {Comb}P. Del'Haye, T. Herr, E. Gavartin, M. L. Gorodetsky, R.
Holzwarth, and T. J. Kippenberg, \textrm{Phys. Rev. Lett.} \textbf{107},
063901 (2011).

\bibitem {Nature2006Kimble}T. Aoki, B. Dayan, E. Wilcut, W. P. Bowen, A. S.
Parkins, T. J. Kippenberg, K. J. Vahala, and H. J. Kimble, \textrm{Nature
(London)} \textbf{443}, 671 (2006).

\bibitem {Nano2006Wang}Y.-S. Park, A. K. Cook, and H. Wang, \textrm{Nano
Lett.} \textbf{6}, 2075 (2006).

\bibitem {XiaoPRARap}Y.-F. Xiao \textit{et al.}, \textrm{Phys. Rev. A
}\textbf{85}, 031805(R) (2012).

\bibitem {PRL2005Vahala}T. J. Kippenberg, H. Rokhsari, T. Carmon, A. Scherer,
and K. J. Vahala, \textrm{Phys. Rev. Lett.} \textbf{95}, 033901 (2005).

\bibitem {Nature2011Painter}A. H. Safavi-Naeini, T. P. Mayer Alegre, J. Chan,
M. Eichenfield, M. Winger, Q. Lin, J. T. Hill, D. E. Chang, and O. Painter,
\textrm{Nature} \textbf{472}, 69 (2011).

\bibitem {Science2012Wang}C. Dong, V. Fiore, M. C. Kuzyk, and Hailin Wang,
\textrm{Science} \textbf{338}, 1609 (2012).

\bibitem {PRA1996Haroche}V. Sandoghdar, F. Treussart, J. Hare, V.
Lefevre-Seguin, J.-M. Raimond, and S. Haroche, \textrm{Phys. Rev. A}
\textbf{54}, R1777 (1996).

\bibitem {Microlasing}G. S. Solomon, M. Pelton, and Y. Yamamoto, \textrm{Phys.
Rev. Lett.} \textbf{86}, 3903 (2001).

\bibitem {APL2003Yang}L. Yang, D. K. Armani, and K. J. Vahala, \textrm{Appl.
Phys. Lett.} \textbf{83}, 825 (2003).

\bibitem {microlaserXiao}Y.-F. Xiao, C.-H. Dong, C.-L. Zou, Z.-F. Han, L.
Yang, and G.-C. Guo, \textrm{Opt. Lett.} \textbf{34}, 509 (2009).

\bibitem {APL2002Arnold}F. Vollmer, D. Braun, A. Libchaber, M. Khoshsima, I.
Teraoka, and S. Arnold, \textrm{Appl. Phys. Lett.} \textbf{80}, 4057 (2002).

\bibitem {OL2006Fan}I. M. White, H. Oveys, and X. Fan, \textrm{Opt. Lett.}
\textbf{31}, 1319 (2006).

\bibitem {PNAS2008Keng}F. Vollmer, S. Arnold, and D. Keng, \textrm{Proc. Natl.
Acad. Sci. U.S.A.} \textbf{105}, 20701 (2008).

\bibitem {NatPho2010Yang}J. Zhu, S. K. Ozdemir, Y.-F. Xiao, L. Li, L. He,
D.-R. Chen, and L. Yang, \textrm{Nature Photon.} \textbf{4}, 46 (2010).

\bibitem {XiaoOE}Y.-F. Xiao, V. Gaddam, and L. Yang, \textrm{Opt. Express}
\textbf{16}, 12538 (2008).

\bibitem {PrismPRA1989}V. B. Braginsky, M. L. Gorodetsky, and V. S. Ilchenko,
\textrm{Phys. Lett. A} \textbf{137}, 393 (1989).

\bibitem {FBOL1997}J. C. Knight, G. Cheung, F. Jacques, and T. A. Birks,
\textrm{Opt. Lett.} \textbf{22}, 1129 (1997).

\bibitem {PRL2000Vahala}M. Cai, O. Painter, and K. J. Vahala, \textrm{Phys.
Rev. Lett.} \textbf{85}, 74 (2000).

\bibitem {FbOL1999}V. S. Ilchenko, X. S. Yao, and L. Maleki, \textrm{Opt.
Lett.} \textbf{24}, 723 (1999).

\bibitem {NonresonantPumping}S.-B. Lee, J. Yang, S. Moon, J.-H. Lee, and K.
An, \textrm{Appl. Phys. Lett.} \textbf{90}, 041106 (2007).

\bibitem {DynamicalAn}J. Yang, S.-B. Lee, S. Moon, S.-Y. Lee, S.W. Kim,
TruongThiAnh Dao, J.-H. Lee, and K. An, \textrm{Phys. Rev. Lett.}
\textbf{104}, 243601 (2010).

\bibitem {Nature}J. U. N\"{o}ckel and A. D. Stone, \textrm{Nature}
\textbf{385}, 45 (1997).

\bibitem {OE2007Wang}Y.-S. Park and H. Wang, \textrm{Opt. Express}
\textbf{15}, 16471 (2007).

\bibitem {Fano}U. Fano, \textrm{Phys. Rev.} \textbf{124}, 1866 (1961).

\bibitem {onpublish}Y.-F. Xiao \textit{et al.}, http://arxiv.org/abs/1209.4441.

\bibitem {Zou}C. L. Zou, F. W. Sun, C. H. Dong, X. W. Wu, J. M. Cui, Y. Yang,
G. C. Guo, Z. F. Han, http://arxiv.org/abs/0908.3531.

\bibitem {Jiang}X.-F. Jiang, Y.-F. Xiao, C.-L. Zou, L. He, C.-H. Dong, B.-B.
Li, Y. Li, F.-W. Sun, L. Yang, and Q. Gong, \textrm{Adv. Mater.} \textbf{24},
OP260 (2012).

\bibitem {KAM}V. I. Arnol'd, \textrm{Russ. Math. Surv.} \textbf{18}, 9 (1963).

\bibitem {OL1994}J. U. N\"{o}ckel, A. D. Stone, and R. K. Chang, \textrm{Opt.
Lett.} \textbf{19}, 1693 (1994).

\bibitem {Science1998}C. Gmachl \textit{et al.}, \textrm{Science}
\textbf{280}, 1556 (1998).

\bibitem {QC}M. Hentschel and K. Richter, \textrm{Phys. Rev. E} \textbf{66},
056207 (2002).

\bibitem {PRAAn}S.-Y. Lee and K. An, \textrm{Phys. Rev. A} \textbf{83}, 023827 (2011).

\bibitem {Husimi}M. Hentschel, H. Schomerus, and R. Schubert,
\textrm{Europhys. Lett.} \textbf{62}, 636 (2003).

\bibitem {WGM}As the regular mode lies on an invariant torus behaving much
like whispering gallery mode, we use WGM to represent it.

\bibitem {Tunnelingrate}A. B\"{a}cker, R. Ketzmerick, S. L\"{o}ck, and L.
Schilling, \textrm{Phys. Rev. Lett} \textbf{100}, 104101 (2008).

\bibitem {QuantumNoise}C.W. Gardiner and P. Zoller, Quantum Noise, 3rd ed.
(Springer, Berlin, 2004).

\bibitem {Fresnel}H. Schomerus and M. Hentschel, \textrm{Phys. Rev. Lett.}
\textbf{96}, 243903 (2006).

\bibitem {TransmissionKorea}J. Yang, S.-B. Lee, S. Moon, S.-Y. Lee, S. W. Kim,
and K. An, \textrm{Opt. Express} \textbf{18}, 26141 (2010).

\bibitem {PRL1986}T. Geisel, G. Radons, and J. Rubner \textrm{Phys. Rev.
Lett.} \textbf{57}, 2883 (1986).

\bibitem {PRL2007}I. I. Rypina, M. G. Brown, F. J. Beron-Vera, H. Kocak, M. J.
Olascoaga, and I. A. Udovydchenkov, \textrm{Phys. Rev. Lett.} \textbf{98},
104102 (2007).

\bibitem {TurnstileAn}J.-B. Shim, S.-B. Lee, S. W. Kim, S.-Y. Lee, J. Yang, S.
Moon, J.-H. Lee, and K. An, \textrm{Phys. Rev. Lett.} \textbf{100}, 174102 (2008).

\bibitem {PRLsong}Q. Song, L. Ge, B. Redding, and H. Cao, \textrm{Phys. Rev.
Lett.} \textbf{108}, 243902 (2012).
\end{thebibliography}
\end{document}